# Threading Through Macrocycles Enhances the Performance of Carbon Nanotubes as Polymer Fillers

*Alejandro López-Moreno,† Belén Nieto-Ortega,† Maria Moffa,‡ Alberto de Juan,† M. Mar Bernal,† Juan P. Fernández-Blázquez,§ Juan J. Vilatela,§ Dario Pisignano*,‡,¥ and Emilio M. Pérez*,†*

† IMDEA Nanoscience, C/Faraday 9, Ciudad Universitaria de Cantoblanco, 28049, Madrid, Spain.

‡ Istituto Nanoscienze-CNR, Euromediterranean Center of Nanomaterial Modelling and Technology (ECMT), via Arnesano, 73100, Lecce, Italy.

§ IMDEA Materials, Eric Kandel 2, Getafe, 28005, Madrid, Spain.

¥ Dipartimento di Matematica e Fisica "Ennio De Giorgi", Università del Salento, via Arnesano, Lecce, Italy

Corresponding Author E-mail: emilio.perez@imdea.org, dario.pisignano@unisalento.it

ABSTRACT

In this work we study the reinforcement of polymers by mechanically interlocked derivatives of single-walled carbon nanotubes (SWNTs). We compare the mechanical properties of fibers made of polymers and of composites with pristine single-walled carbon nanotubes (SWNTs), mechanically interlocked derivatives of SWNTs (MINTs) and the corresponding supramolecular models. Improvements of both Young's modulus and tensile strength of up to 200% were observed for the polystyrene-MINTs samples with an optimized loading of just 0.01 wt.%, while the supramolecular models with identical chemical composition and loading showed negligible or even detrimental influence. This behavior is found for three different types of SWNTs and two types of macrocycles. Molecular dynamics simulations show that the polymer adopts an elongated conformation parallel to the SWNT when interacting with MINT fillers, irrespective of the macrocycle chemical nature, whereas a more globular structure is taken upon facing with either pristine SWNTs or supramolecular models. The MINT composite architecture thus leads to a more efficient exploitation of the axial properties of the SWNTs and of the polymer chain at the interface, in agreement with experimental results. Our findings demonstrate that the mechanical bond imparts distinctive advantageous properties to SWNT derivatives as polymer fillers.

Carbon nanotubes are extensively used as reinforcing fillers in composites due to their extraordinary mechanical and structural properties. Since the report in this field by Ajayan *et al.*,[1] several materials where the mechanical and/or electrical properties of polymers have been significantly improved through nanotubes fillers have been demonstrated, and used for different applications,[2-16] including improved batteries, mechanically reinforced materials[17, 18] and sensors.[19,20]

To fully exploit the properties of single wall carbon nanotubes (SWNTs) as fillers in polymer matrices, a lot of research has been addressed to their chemical modification. In this framework, the mechanical bond is very attractive due to its dynamic features,[21,22] which have allowed for the construction of artificial molecular machines.[23-30] The mechanical bond is also very relevant for polymer science: polyrotaxanes, polycatenanes, and supramolecular polymers including mechanically interlocked molecules have all been investigated.[31-39] The reinforcement effect of B/SiOx nanocomposites through the formation of interlocked "necklaces" has also been described.[40, 41]

The mechanical link was introduced by some of us as a tool for the chemical manipulation of SWNTs very recently.[42-45] We used a U-shaped precursor featuring two units of a recognition element for SWNTs connected through an aromatic spacer, and further decorated with alkene-terminated alkyl spacers of different lengths. Using pyrene and π-extended derivatives of tetrathiafulvalene, both of which have high affinity for SWNTs,[46-49] We could template the ring closing metathesis (RCM) of the U-shaped precursor around the nanotubes, forming mechanically interlocked derivatives of SWNTs (MINTs, Figure 1a). Thanks to the extreme aspect ratio of the nanotubes, which prevents dissociation of the macrocycles from the nanotubes once they are formed around them, MINTs showed stability comparable to that of covalently modified nanotubes, while maintaining the native structure of the SWNTs. Since rotaxanes and pseudorotaxanes are both topologically identical,[50, 51] and the major difference between them is their kinetic stability,[52] we believe our MINT derivatives can be considered mechanically interlocked, despite the lack of explicit stoppers.

To effectively transfer the anisotropic properties of elongated fillers such as SWNTs to composites, a parallel orientation in the matrix and a strong interaction with the polymer are required. In principle, the parallel orientation along the prevalent direction of macromolecular chains can be favored by electrospinning, due to the very high elongational strain rates applied,[53,54,55] while the noncovalent interactions between polymer and filler can be tuned chemically. Recently, the groups led by Pisignano and Credi have described that various dynamic properties of rotaxane-type molecules are conserved within electrospun fibers.[56]

Here, we present our results on the influence of the mechanical bond on the mechanical properties of SWNT-based nanocomposites. We incorporate MINTs in polystyrene fibers and study their tensile properties. The merits of the MINT functionalization approach manifest as substantial enhancements in Young's modulus and tensile strength. In comparison, non-interlocked model samples of identical chemical composition show no positive effect.

RESULTS AND DISCUSSION

We utilized two types of macrocycles (Figure 1b) and three types of SWNTs of different diameter, length and electronic character. In particular, we used pyrene (**1**) and exTTF (**2**) based macrocycles and (6,5)-enriched nanotubes purchased from Sigma Aldrich Co. (0.7-0.9 nm in diameter, length ≥ 700 nm, mostly semiconducting, 95% purity) denoted as (6,5)-SWNTs, plasma-purified SWNTs (pp-SWNTs) purchased from Cheap Tubes Inc. (0.8-1.6 nm in diameter, length 3-30 μm, mostly metallic, 99 % purity), and shorter COOH functionalized SWNTs (o-SWNTs) purchased from Cheap Tubes Inc. (0.8-1.6 nm in diameter, length 0.5-2.0 μm, mostly metallic, 99 % purity). These various types of samples allowed us to discriminate mechanical reinforcement arising from differences in SWNT length or dispersion quality from those directly due to the MINT functionalization.

The general method for the synthesis of MINTs has been reported elsewhere.[42-45] Briefly, we use a clipping strategy in which a suspension of SWNTs is treated with the adequate bis-alkene U-shape precursor and Grubb's $2^{nd}$ generation catalyst. After supramolecular association of the U-shape, it can be closed around the SWNT to form MINTs. Non-interlocked macrocycles and U-shapes, oligomers, catalyst, and all other byproducts are removed by extensive washes with dichloromethane. The interlocked macrocycles stay in place without the need for "stoppers" due to the extreme aspect ratio of the SWNTs. All samples used in this study were adequately characterized by standard methods, including TGA, Raman, UV-vis-NIR and TEM. Figure 1 shows representative examples of TGA curves, Raman spectra and TEM micrographs. TGA evidences that, following MINT-forming reaction, the SWNTs showed organic functionalization between 27 and 43%, keeping stable even after reflux in tetrachloroethane for 30 min. No major shifts and no increase in the $I_D/I_G$ ratio upon functionalization were found in the Raman spectra, confirming that the functionalization is noncovalent. HRTEM allows visualizing individual macrocycles around the SWNTs in the MINT samples (for comprehensive characterization, including control experiments, please see the Supporting Information and Refs. 42-45).

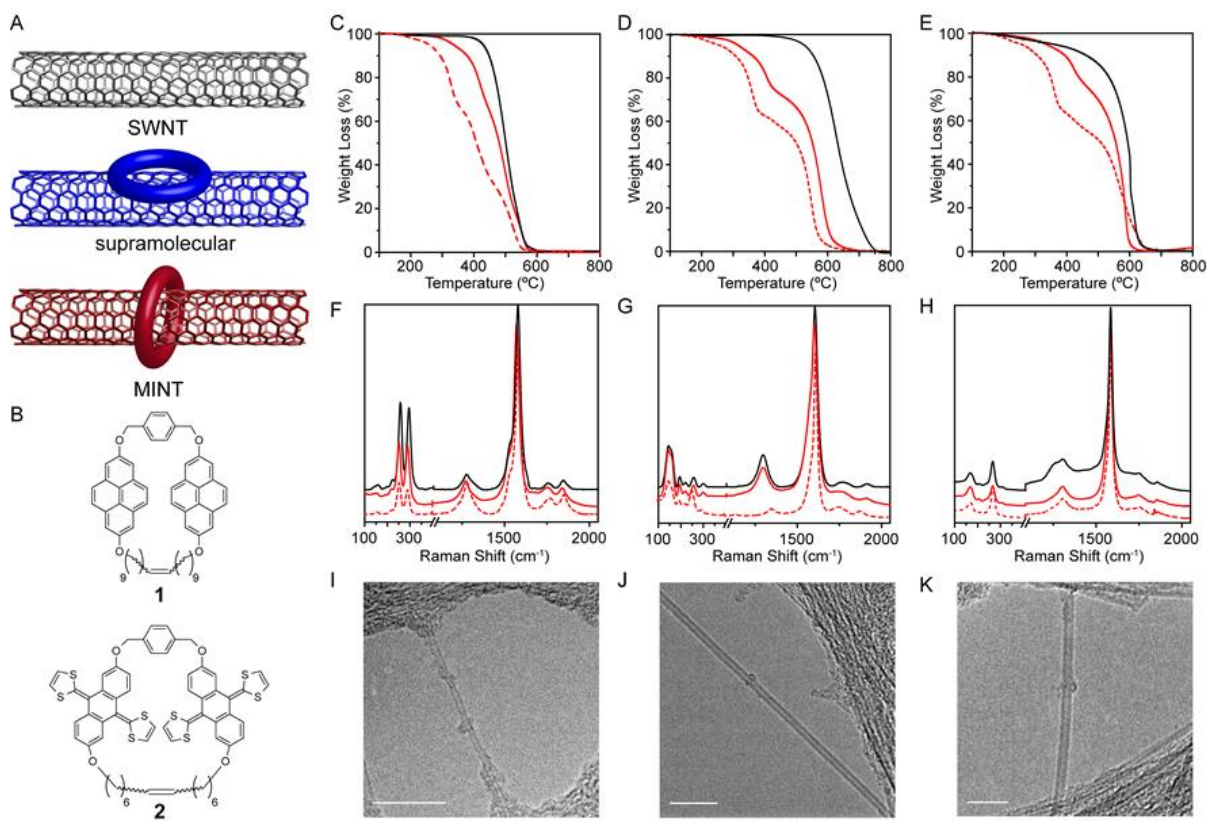

**Figure 1.** a) Schematic representation of the three fillers investigated: SWNTs, supramolecular associates and MINTs; b) Chemical structure of macrocycles **1** and **2**. c-k) Characterization of MINT derivatives. TGA analysis of c) pristine (6,5)-SWNTs, (black) MINT$_{(6,5)}$-**1** (red) and MINT$_{(6,5)}$-**2** (dashed red); d) pristine pp-SWNTs (black), MINT$_{(pp)}$-**1** (red) and MINT$_{(pp)}$-**2** (dashed red); e) pristine o-SWNTs (black), MINT$_{(o)}$-**1** (red) and MINT$_{(o)}$-**2** (dashed red); Raman spectra of f) (6,5)-SWNTs (black), MINT$_{(6,5)}$-**1** (red) and MINT$_{(6,5)}$-**2** (dashed red); g) pp-SWNTs (black), MINT$_{(pp)}$-**1** (red) and MINT$_{(pp)}$-**2** (dashed red); h) o-SWNTs (black), MINT$_{(o)}$-**1** (red) and MINT$_{(o)}$-**2** (dashed red); TEM images of nanotubes (showing macrocycles around nanotubes) in i) MINT$_{(6,5)}$-**2**; j) MINT$_{(pp)}$-**2** and k) MINT$_{(o)}$-**2**. Scale bars are 10 nm. TGAs were run in air at a heating rate of 10 ºC min$^{-1}$. All Raman spectra are the average of ten different measurements at $\lambda_{exc}$ = 785 nm.

We prepared suspensions of the SWNT derivatives through ultrasonication. To avoid the presence of aggregates that could affect the mechanical properties, the suspensions were centrifuged and then polystyrene was added. Electrospinning was carried out using a commercially available system, operating with an applied inter-electrode bias of 14 kV and a flow rate of 1 mL h$^{-1}$. Experimental details and videos of the electrospinning process can be found in the Supporting Information. The filler loading was optimized to 0.01 wt% respect to polystyrene, since larger loadings lead to defective fibers (Supporting Information). With this loading, fibers showed seamless and uniform surfaces, without discernable beads or nanotube aggregates. Figure 2 displays typical scanning electron (SEM) micrographs of fibers made of pristine polystyrene, and of those with o-SWNTs-based fillers, as representative examples (other samples are shown in Supporting Information).

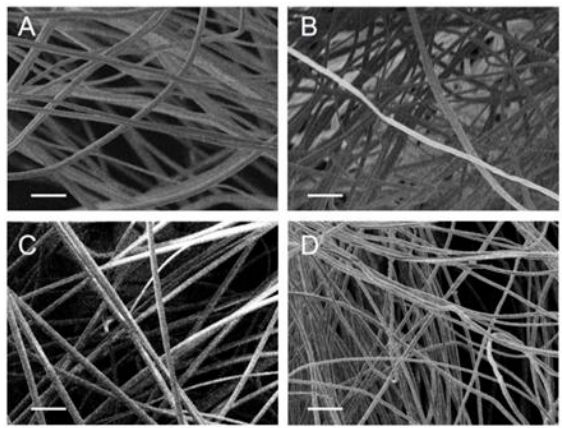

**Figure 2.** SEM images of: a) Polystyrene fibers; b) o-SWNTs; c) MINT$_{(o)}$-**1**; d) o-SWNTs·**1**. Inset scale: 10 μm.

The diameter distribution of the fibers is within the same range of 1.3-1.8 μm for all samples (Figure S7). Pristine polystyrene fibers are slightly thicker (2.2 ± 0.6 μm) as expected because of the lower solution conductivity.[57] The (6,5)-SWNTs and pp-SWNTs samples are similar to o-SWNTs in structure and size as shown Figure S8. The mechanical properties of the fibers were then determined using a dynamic mechanical analyzer (DMA Q800, TA Instruments). Each nanocomposite (n = 3 specimens) was cut into 1×4 cm2 pieces to define samples with thickness 0.15-0.18 mm. All samples had comparable area density (ca. 2.5 mg cm$^{-2}$) and thus the test specimens had similar linear densities too thus ensuring stable force to stress normalization in measurements. Force-displacement curves were recorded at 1 N min$^{-1}$ (up to 18 N).

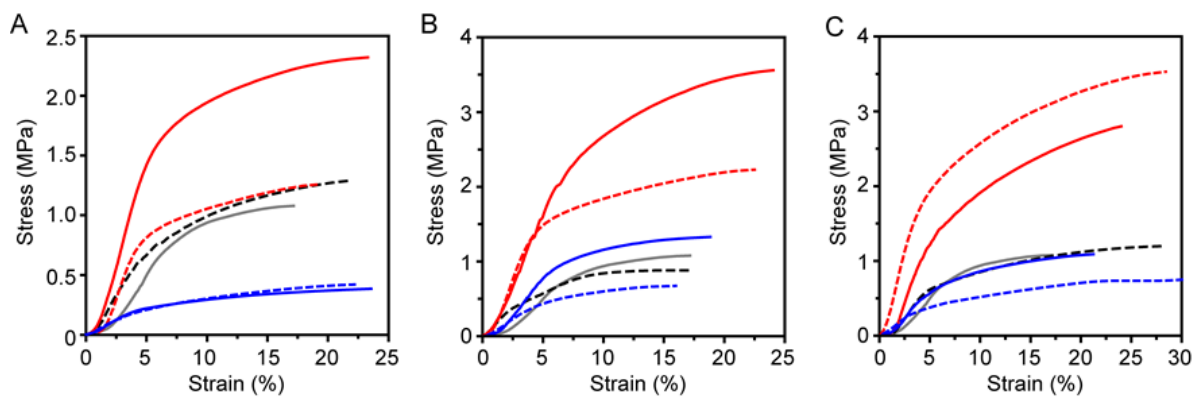

**Figure 3.** Representative stress/strain curves of a) polystyrene (grey), and its composites with (6,5)-SWNTs (dashed black), MINT$_{(6,5)}$-**1** (red), (6,5)-SWNTs·**1** (blue), MINT$_{(6,5)}$-**2** (dashed red) and (6,5)-SWNTs·**2** (dashed blue); b) polystyrene (grey), and its composites with pp-SWNTs (dashed black), MINT$_{(pp)}$-**1** (red), pp-SWNTs·**1** (blue), MINT$_{(pp)}$-**2** (dashed red) and pp-SWNTs·**2** (dashed blue); c) polystyrene (grey), and its composites with o-SWNTs (dashed black), MINT$_{(o)}$-**1** (red), o-SWNTs·**1** (blue), MINT$_{(o)}$-**2** (dashed red) and o-SWNTs·**2** (dashed blue).

Figure 3 displays stress/strain curves for reference polystyrene (grey) and the nanocomposites explored: with pristine SWNTs (black), SWNTs + macrocycle supramolecular complex (blue), and MINT (red) for all types of SWNTs. The MINT samples present substantially higher modulus, yield and tensile strengths than all control samples. Interestingly, in the supramolecular systems the macrocycle reduces dramatically both modulus and strength, suggesting that it acts as a plasticizer that weakens the SWNT/matrix interface. The traditional composite has similar tensile properties to the pure polystyrene matrix, including ductility. The implication is that at this low volume fraction even pure SWNTs are well dispersed, for otherwise in aggregated form they would most likely act as defects that would reduce ductility. This supports the view that the improvement in mechanical reinforcement obtained using the MINT strategy is due to a more efficient stress transfer across the SWNT/polymer interface (*vide infra*). The Young's moduli and tensile strengths of all samples are displayed in Figure 4 and Table S1.

General trends are clearly evidenced. Firstly, the mechanical properties of fibers are only slightly reinforced by pristine SWNTs fillers. Secondly, the use of MINTs leads instead to a significant improvement of both the Young's modulus and the tensile strength in all samples, irrespective of the type of nanotube or macrocycle. Lastly, the supramolecular fillers have negligible or even detrimental effects on the mechanical properties of the polystyrene fibers. For instance, the samples in which pristine (6,5)-SWNTs were used as fillers showed a Young's modulus of $(18 \pm 1)$ MPa and a tensile strength of $(1.26 \pm 0.06)$ MPa, whereas the pure polystyrene fibers showed $(15 \pm 1)$ MPa and $(1.09 \pm 0.03)$ MPa, respectively. In contrast, the $MINT_{(6,5)}$-1 samples showed $(32 \pm 6)$ MPa and $(2.0 \pm 0.3)$ MPa, that is, a remarkable improvement of 110 % in the Young's modulus and of 80 % in the tensile strength. Meanwhile, the supramolecular filler (6,5)-SWNTs·1 yielded $(7 \pm 1)$ MPa and $(0.39 \pm 0.03)$ MPa as Young's modulus and tensile strength, respectively, which implies a variation of −53 % in the Young's modulus and of −64 % in the tensile strength with respect to the pristine polymer.

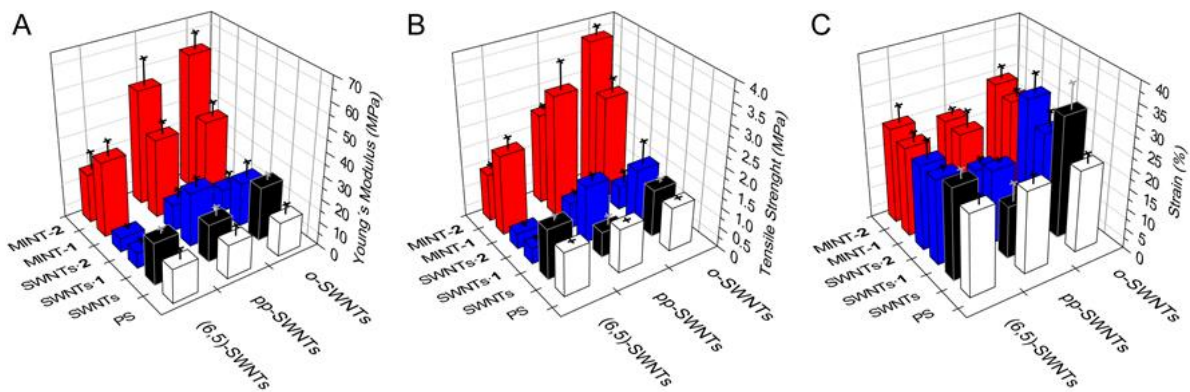

**Figure 4.** a) Young's modulus of polystyrene (white), SWNTs (black), supramolecular complexes (blue) and MINTs (red) with (6,5)-SWNTs (left), pp-SWNTs (center) and o-SWNTs (right); b) Tensile strength of polystyrene (white), SWNTs (black), supramolecular complexes (blue) and MINTs (red) with (6,5)-SWNTs (left), pp-SWNTs (center) and o-SWNTs (right); c) Strain of polystyrene (white), SWNTs (black), supramolecular complexes (blue) and MINTs (red) with (6,5)-SWNTs (left), pp-SWNTs (center) and o-SWNTs (right).

The trends for macrocycle 2 are identical, although with quantitatively smaller effects. In the case of pristine pp-SWNTs, we observed no significant variation in the Young's modulus and a decrease of −50 % in the tensile strength with respect to polystyrene. Meanwhile, the MINT$_{(pp)}$-1 and MINT$_{(pp)}$-2 fillers showed an increase of 130 % and 230 % in the Young's modulus and 170 % and 106 % in the tensile strength, respectively. The supramolecular models showed very small improvements in the case of macrocycle 1 and slightly detrimental effects for macrocycle 2.

Finally, for the pristine o-SWNTs filled samples, the variation in Young's modulus with respect to polystyrene is of 53 % and of only 8 % in tensile strength. Again, the mechanically interlocked samples lead to a well-defined improvement, offering 130 % and 290 % variations in Young's modulus, and 140 % and 240 % increase in tensile strength for MINT(o)-1 and MINT(o)-2, respectively. Just like with the other types of nanotubes, the supramolecular fillers offered no improvements in the mechanical properties over pure polystyrene.

No significant changes were observed in the strain-to-break among samples with the same kind of nanotubes (Figure 4c).

Complex effects could be responsible for the improvement observed in the mechanical properties of the composites, including nanoscale friction at the polymer-nanocarbon interface.[58,59] In order to gain atomic understanding of our system, molecular dynamics (MD) simulations were performed using the AMBER force field,[60] which accounts for dispersion interactions. To mimic our experimental conditions as much as possible, MD calculations were carried out using a (6,5)-SWNT of 400 atoms, to ensure the same SWNT/macrocycle ratio measured experimentally. The polystyrene fiber consisted of 36 residues, which were introduced in a fully extended conformation to emulate the electrospinning conditions. Initial configuration of the composites and computational details are described in the Supporting Information. Figure 5 shows the equilibrated structures of polystyrene, and its composites with SWNTs, SWNT·1, SWNT·2, MINT-1 and MINT-2. Due to the flexible backbone, after 0.4 ns a highly-twisted, globular structure is adopted to maximize intramolecular interactions (Figure 5a). A similar picture dominates the first frames of the simulations with the nanotube fillers, until polymer-nanotube intermolecular interactions become relevant. Upon stabilization (after approximately 2 ns of simulation time, see the Supporting Information) we observe very clear differences between the various fillers. The pristine nanotubes allow polystyrene to adopt a globular structure, very similar to that found for pure polystyrene (Figure 5b). In the supramolecular controls, the fiber tries to maximize short contacts with both macrocycle and SWNT, which results in a slightly more distorted structure (Figures 5c and 5d). Finally, in the MINT samples the positioning of the macrocycles around the nanotubes results in less surface available for interaction with the polystyrene fiber, which reacts by adopting a significantly more extended conformation in order to maximize noncovalent interactions with the SWNT (Figures 5e and 5f). As a quantitative metric for these observations, we measured the dihedral angles of the polystyrene backbone for each case for a total of 500 frames in the last nanosecond of our simulation (Figure 5g). An average of around 130º is found for the MINTs, compared to an average of 95º for the supramolecular compounds and approximately 105º for the polymer-SWNT model. The extended conformation of polystyrene according to MD simulations is more pronounced in the case of MINT-1 when compared to MINT-2, while the results for the supramolecular models are fundamentally independent of the structure of the macrocycle, in direct correlation with experimental results. Polymer chain extension and orientation are established prerequisites to produce strong/stiff polymeric materials, for example in the form of high-performance fibres.[61,62] The MINT-induced polymer conformation extracted from MD simulations is in line with such arrangement and agrees with the higher degree of reinforcement observed for MINT-containing composites.

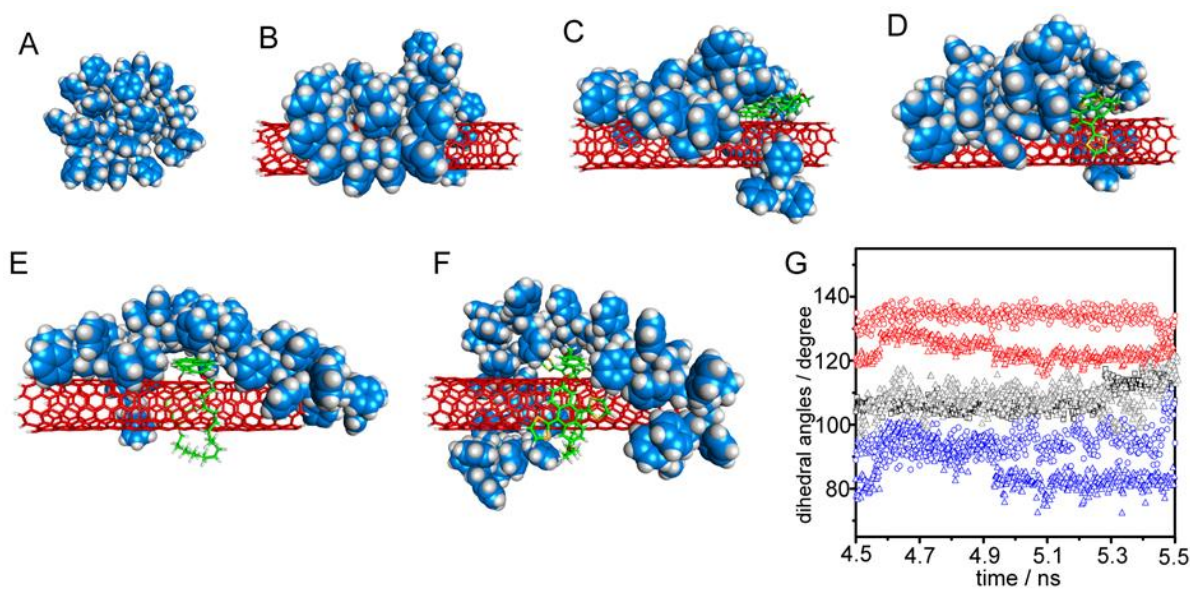

**Figure 5**. MD snapshots of a) Polystyrene, and its composites with b) SWNTs, c) SWNT·**1**, d) SWNT·**2**, e) MINT-**1**, and f) MINT-**2** after MD simulations. Carbon atoms are shown in red for the SWNTs, green for the macrocycles and cyan for polystyrene. Hydrogens are shown in white, oxygen in red and sulfur in yellow. g) Average dihedral angles of the polystyrene backbone for the last nanosecond of the MD simulation. Color code: polystyrene (grey triangle), and its composites with SWNTs (black square), SWNT·**1** (blue circle), SWNT·**2** (blue triangle), MINT-**1** (red circle), and MINT-**2** (red triangle).

CONCLUSIONS

In summary, we have shown that mechanical interlocking is a strategy to optimize the performance of SWNT fillers with regard to their mechanical properties. Very low loading of 0.01% results in improvement of the Young's modulus and tensile strength of the fibers of over 200%. In comparison, fillers with identical chemical composition but lacking the interlocked architectures showed negligible or even detrimental effects. Moreover, by using up to three different kinds of nanotubes, two macrocycles, and the corresponding supramolecular controls, we have demonstrated that the positive effect is general to the MINTs samples, as the trends hold in all cases under study. MD simulations show that this effect originates from a superior ability of the MINT samples to induce extended conformation in the polystyrene fibers, which allows for an optimized transfer of stress between matrix and SWNTs.

METHODS/EXPERIMENTAL

(6,5)-enriched nanotubes were purchased from Sigma Aldrich Co. (0.7-0.9 nm in diameter, length $\geq$ 700 nm, mostly semiconducting, 95% purity), plasma-purified SWNTs (pp-SWNTs) were purchased from Cheap Tubes Inc. (0.8-1.6 nm in diameter, length 3-30 µm, mostly metallic, 99 % purity), and COOH functionalized SWNTs (o-SWNTs) were purchased from Cheap Tubes Inc. (0.8-1.6 nm in diameter, length 0.5-2.0 µm, mostly metallic, 99 % purity). Electrospinning was carried out using a commercially available Spraybase Electrospinning. Thermogravimetric analysis (TGA) was performed using a TA Instruments TGAQ500 with a ramp of 10 °C/min under air from 100 to 1000 °C. Scanning electron microscopy (SEM) micrographs were obtained in a Zeiss EVO HD15 operating at 5 kV. Ultraviolet-visible spectra was obtained in a Varian Cary 50 UV-Vis. Mechanical properties were determined using a dynamic mechanical analyzer (DMA Q800, TA Instruments). Each fibers sample (n = 3 specimens) was cut in 1cm x 4cm rectangular shapes with thickness between 0.15 and 0.18 mm. Stress−strain curves were recorded at a rate of 1 N min−1 (up to 18N).

Linear receptors and MINTs were synthesized as described in references 42-45. The nanotubes (10 mg) were suspended in 10 mL of tetrachloroethane (TCE) through sonication (10 min.) and mixed with linear precursors 1 and 2 (0.01 mmol), and Grubb's 2nd generation catalyst at room temperature for 72 hours. After this time, the suspension was filtered through a PTFE membrane of 0.2 μm pore size, and the solid washed profusely with dichloromethane (DCM). The solid was re-suspended in 10 mL of DCM through sonication for 10 min. and filtered through a PTFE membrane of 0.2 μm pore size again. This washing procedure was repeated three times.

Composites were prepared by direct suspension of MINTs or pristine nanotubes in dimethylformamide (DMF) by sonication at 20ºC for 12 h and the suspensions were centrifuged at 13150 G for 15 minutes to obtain stable suspensions., following the addition of polystyrene (Mw average 350000) 30 % (w/w) and stirring for 12 hours. In the case of supramolecular samples, pristine nanotubes were suspended in the same conditions and preformed macrocycles were added before polystyrene and the mixture was stirred for 12 h. Concentrations of SWNTs, MINTs, and supramolecular models were matched using UV-Vis spectra at 450 nm of the suspension obtained. The prepared solutions were added to a syringe and pumped at 1 mL h$^{-1}$ with a voltage of 14 kV and constant temperature and humidity. All samples were electrospun over a 10 cm diameter round collector to obtain randomly aligned fibers.

*Supporting Information*. Characterization not shown in the main text and computational details. This material is available free of charge *via* the Internet at http://pubs.acs.org.

ACKNOWLEDGMENT

Dedicated to Prof. Nazario Martín, on the occasion of his 60th birthday. The research leading to these results has received funding from the European Research Council under the European Union's Seventh Framework Programme (FP/2007-2013)/ERC Grant Agreements n. 306357 (ERC Starting Grant "NANO-JETS"), n. 307609 (ERC Starting Grant "MINT"), MINECO

(CTQ2014-60541-P, MAT2015-62584-ERC and RyC-2014-15115, Spain) and MAD2D project (S2013/MIT-3007, Comunidad de Madrid). The computational work was supported by the Campus of International Excellence (CEI) UAM+CSIC. Additionally, the authors would like to express their gratitude to the Supercomputing and Bioinnovation Center (SCBI) of the University of Málaga (Spain) for their support and resources.

# SUPPORTING INFORMATION

# Threading Through Macrocycles Enhances the Performance of Carbon Nanotubes as Polymer Fillers

*Alejandro López-Moreno,[†] Belén Nieto-Ortega,[†] Maria Moffa,[‡] Alberto de Juan,[†] M. Mar Bernal,[†] Juan P. Fernández-Blázquez,[§] Juan J. Vilatela,[§] Dario Pisignano\*[‡,¥] and Emilio M. Pérez\*[†]*

† IMDEA Nanoscience, C/Faraday 9, Ciudad Universitaria de Cantoblanco, 28049, Madrid, Spain.

‡ Istituto Nanoscienze-CNR, Euromediterranean Center of Nanomaterial Modelling and Technology (ECMT), via Arnesano, 73100, Lecce, Italy.

§ IMDEA Materials, Eric Kandel 2, Getafe, 28005, Madrid, Spain.

¥ Dipartimento di Matematica e Fisica "Ennio De Giorgi", Università del Salento, via Arnesano, Lecce, Italy.

**Characterization of MINTs samples**

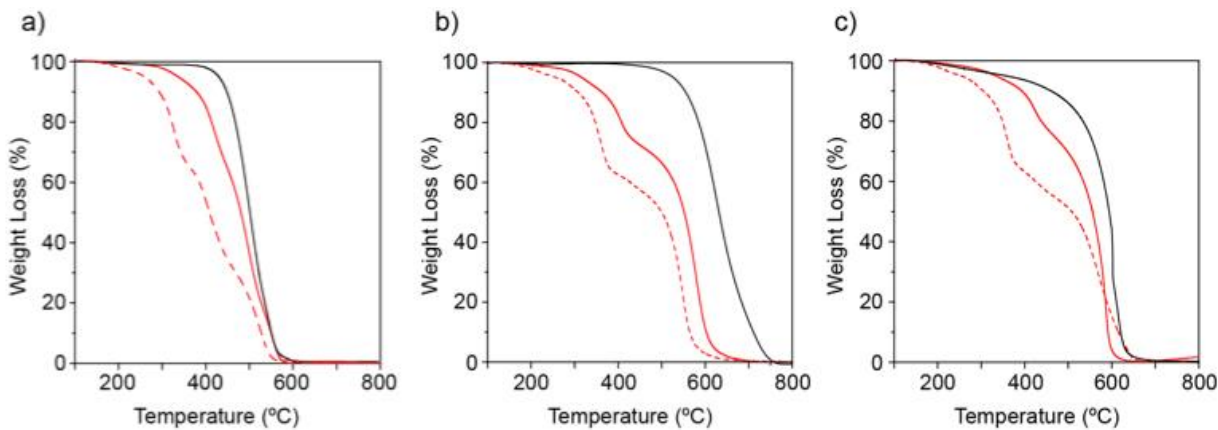

**Figure S1.** TGA analysis of: a) pristine (6,5)-SWNTs (black), MINT$_{(6,5)}$-**1** (red) and MINT$_{(6,5)}$-**2** (dashed red); b) pristine pp-SWNTs (black), MINT$_{(pp)}$-**1** (red) and MINT$_{(pp)}$-**2** (dashed red). TGAs were run in air at a heating rate of 10 ºC min$^{-1}$.

TGA of the solid thus obtained showed weight losses of 33% and 41% for MINT$_{(6,5)}$-**1** and MINT$_{(6,5)}$-**2**, 25% and 35% for MINT$_{(pp)}$-**1** and MINT$_{(pp)}$-**2** at approximately 400°C and 26% and 36% for MINT$_{(pp)}$-**1** and MINT$_{(pp)}$-**2**.

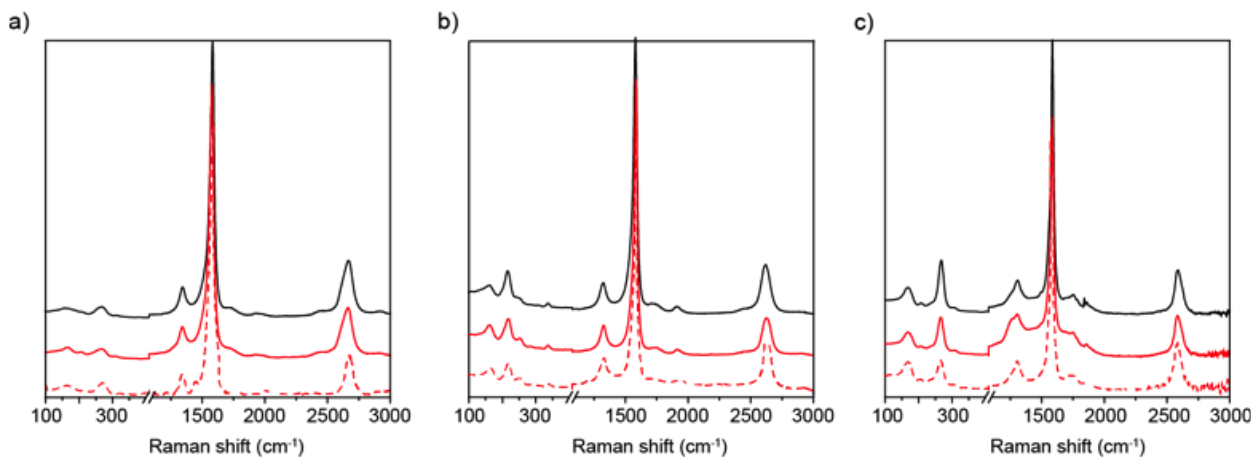

**Figure S2**. Raman spectra of o-SWNTs (black), MINT$_{(o)}$-**1** (red) and MINT$_{(o)}$-**2** (dashed red): a) $\lambda_{exc}$ = 532 nm; b) $\lambda_{exc}$ = 633 nm and c) $\lambda_{exc}$ = 785 nm All spectra are the average of ten different measurements.

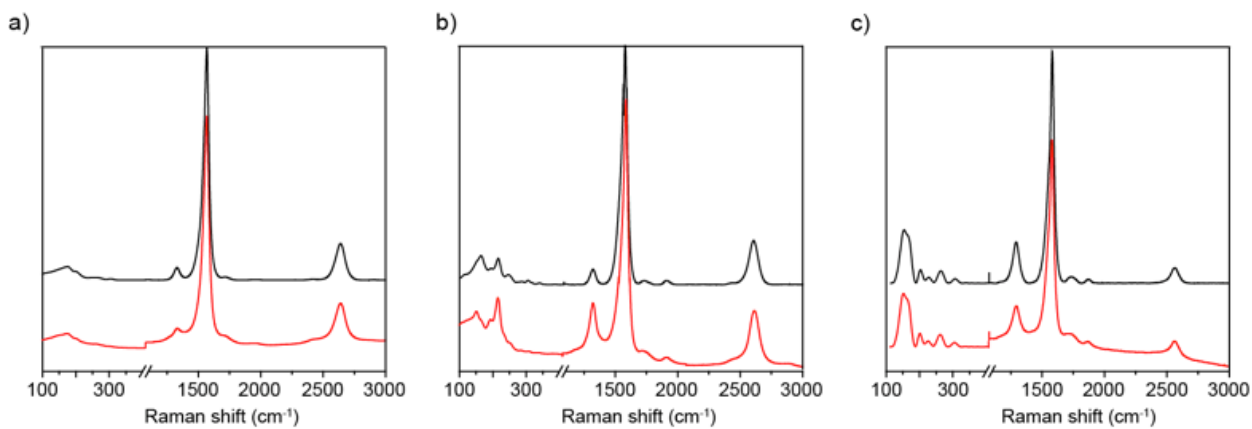

**Figure S3**. Raman spectra of pp-SWNTs (black) and MINT$_{(pp)}$-**1** (red): a) $\lambda_{exc}$ = 532 nm; b) $\lambda_{exc}$ = 633 nm and c) $\lambda_{exc}$ = 785 nm All spectra are the average of ten different measurements.

Raman spectroscopy (Fig. S2 and S3, $\lambda_{exc}$ = 532, 633, and 785 nm) reveals no changes in the spectra with respect to pristine pp-SWNTs and o-SWNTs, as expected for the noncovalent functionalization of SWNTs. In particular, we observed no significant increase in the $I_D/I_G$ ratio and no modification in the RBM intensity, which confirmed that there is no covalent modification of the SWNTs.

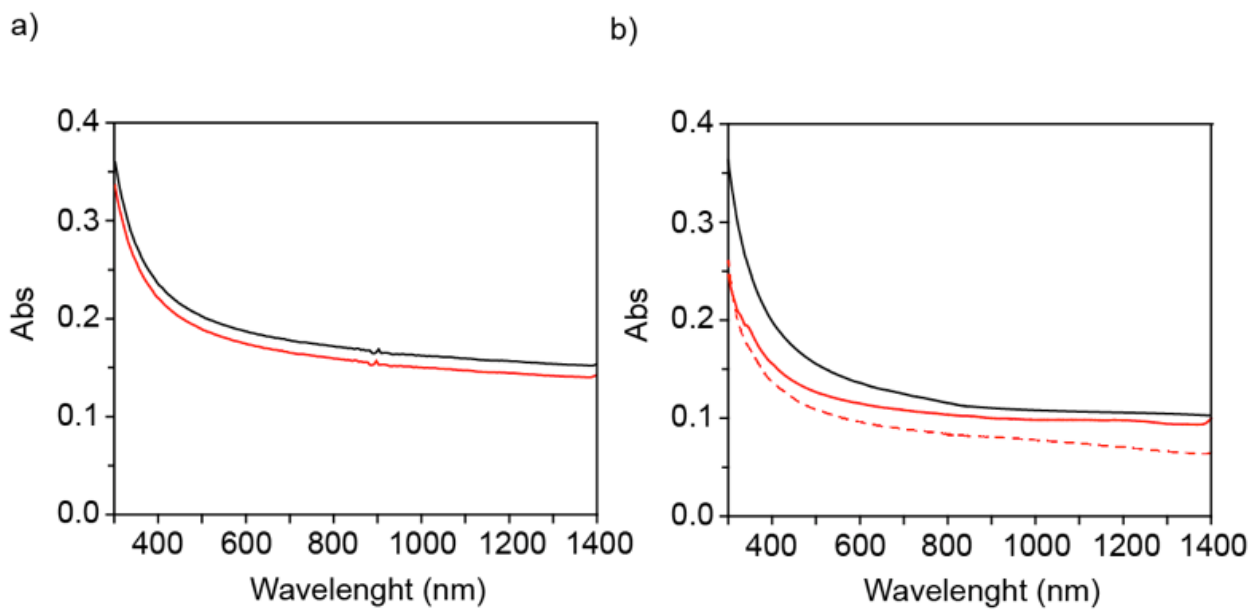

**Figure S4.** UV/Vis spectra (D$_2$O, 1% sodium dodecyl sulfate (SDS) , 298 K) of a) pristine pp-SWNTs (black) and MINT$_{(pp)}$-**1** (red); b) o-SWNTs (black), MINT$_{(o)}$-**1** (red) and MINT$_{(o)}$-**2** (dashed red).

In the absorption spectra (D$_2$O, 1% sodium dodecyl sulphate, 298 K, Fig. S4), the UV region is dominated by the nanotube absorption in both samples, and the characteristic absorption of pyrenes and exTTF in the 300–350 nm and 300-450 nm range respectively is not distinguishable, save for an increase in the relative absorption in this region.

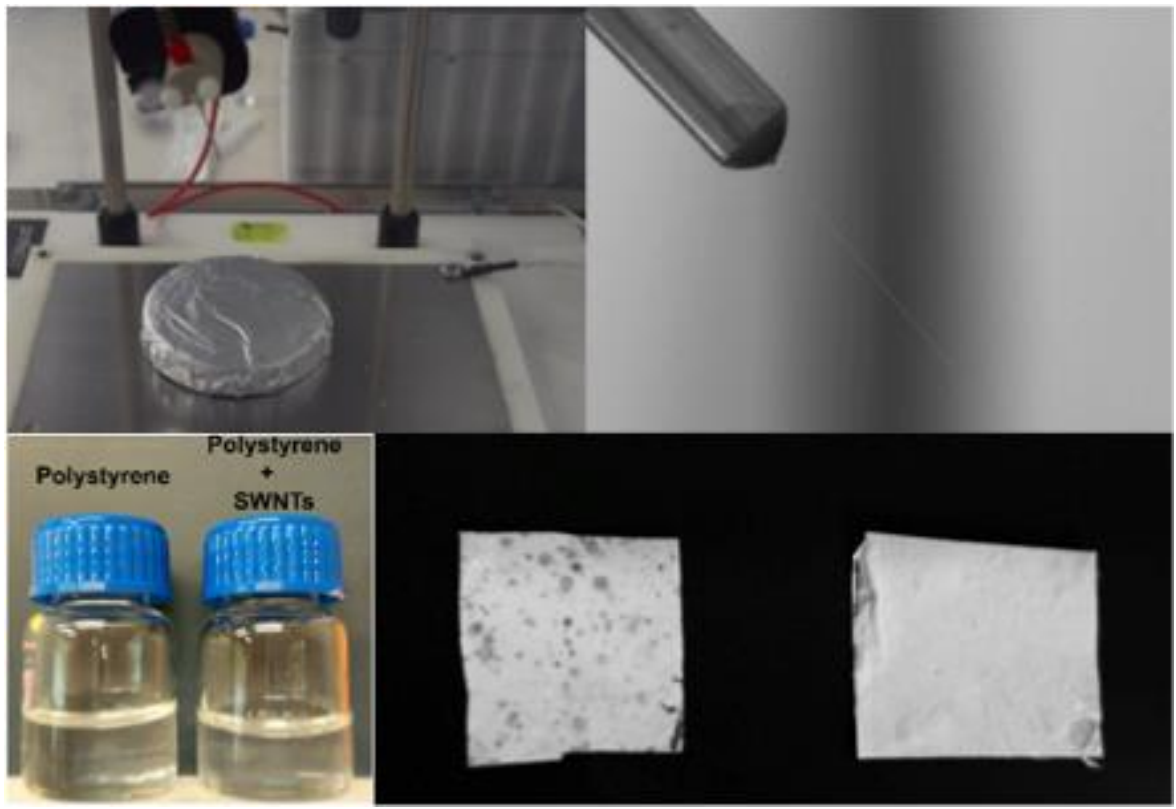

**Figure S5.** Top Left: Electrospinning setup. Top right: Fiber formation in the needle. Bottom left: Polystyrene and polystyrene/ SWNTs solutions in DMF. Bottom: Defective samples of fibers due to a high concentration of SWNTs (left) and a low fiber density.

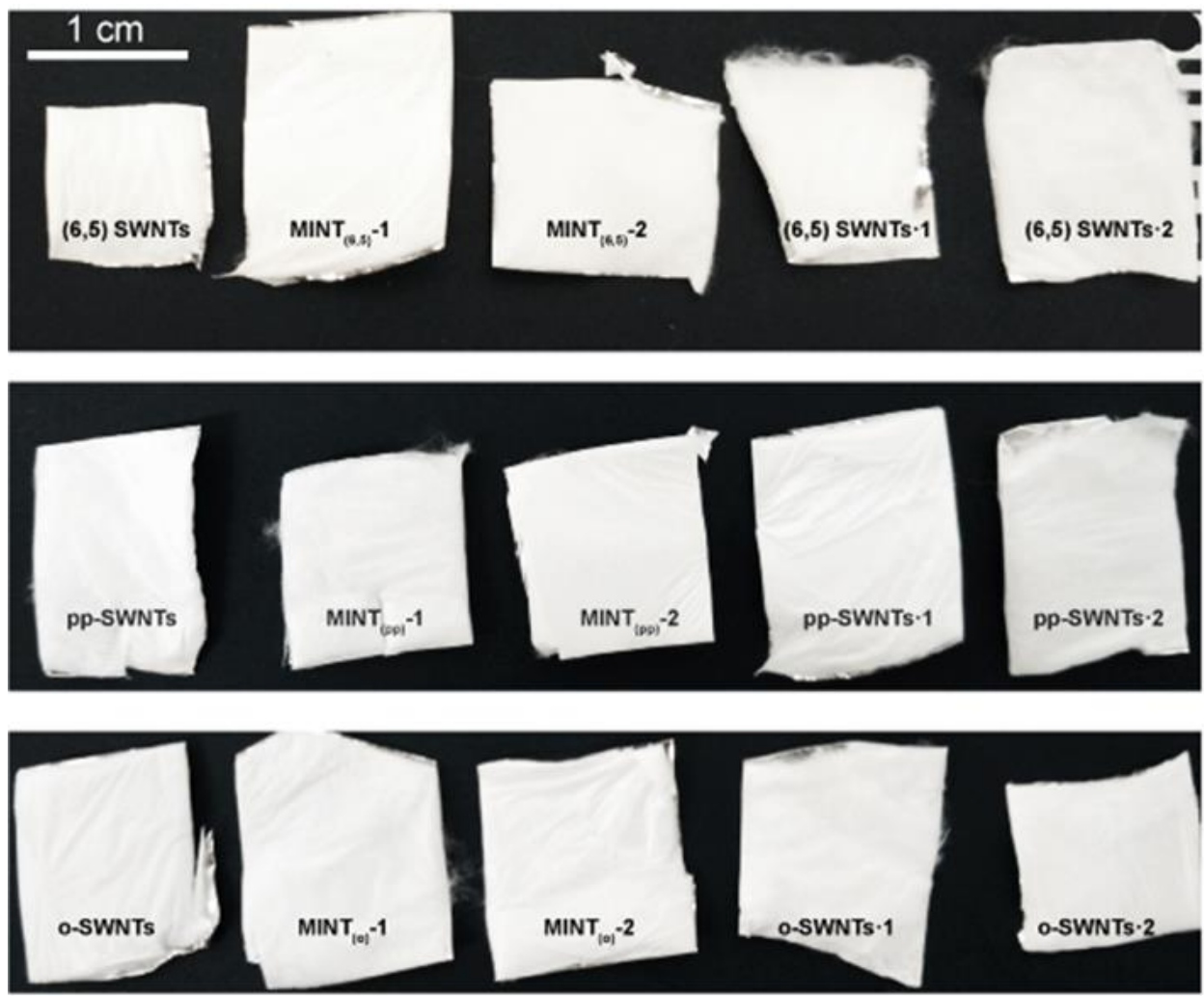

**Figure S6.** Samples of electrospun fibers. No macroscopic differences are observed.

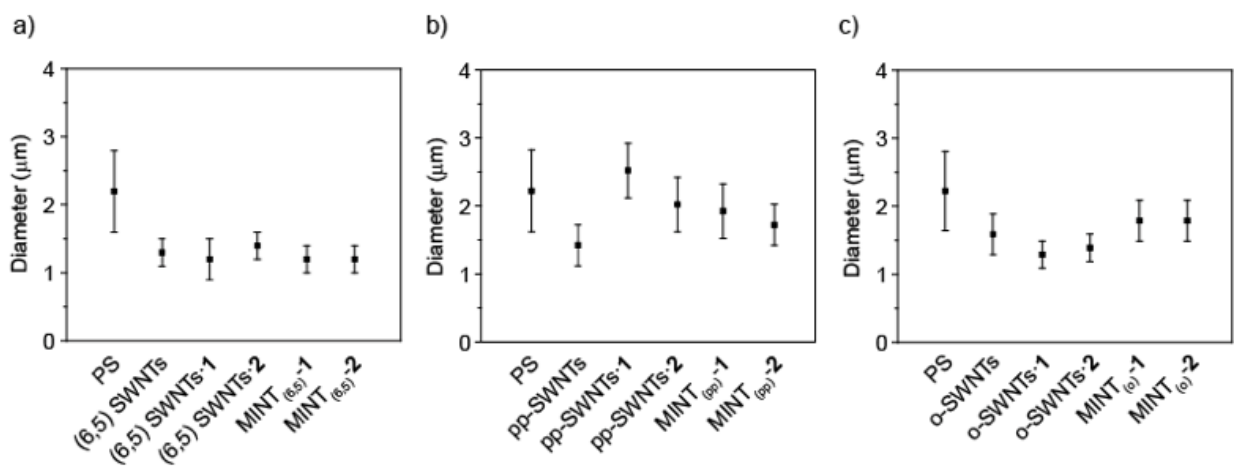

**Figure S7.** Diameter distribution of: a) (6,5) SWNTs samples; b) pp-SWNTs samples and c) o-SWNTs. Mean of 100 measurements.

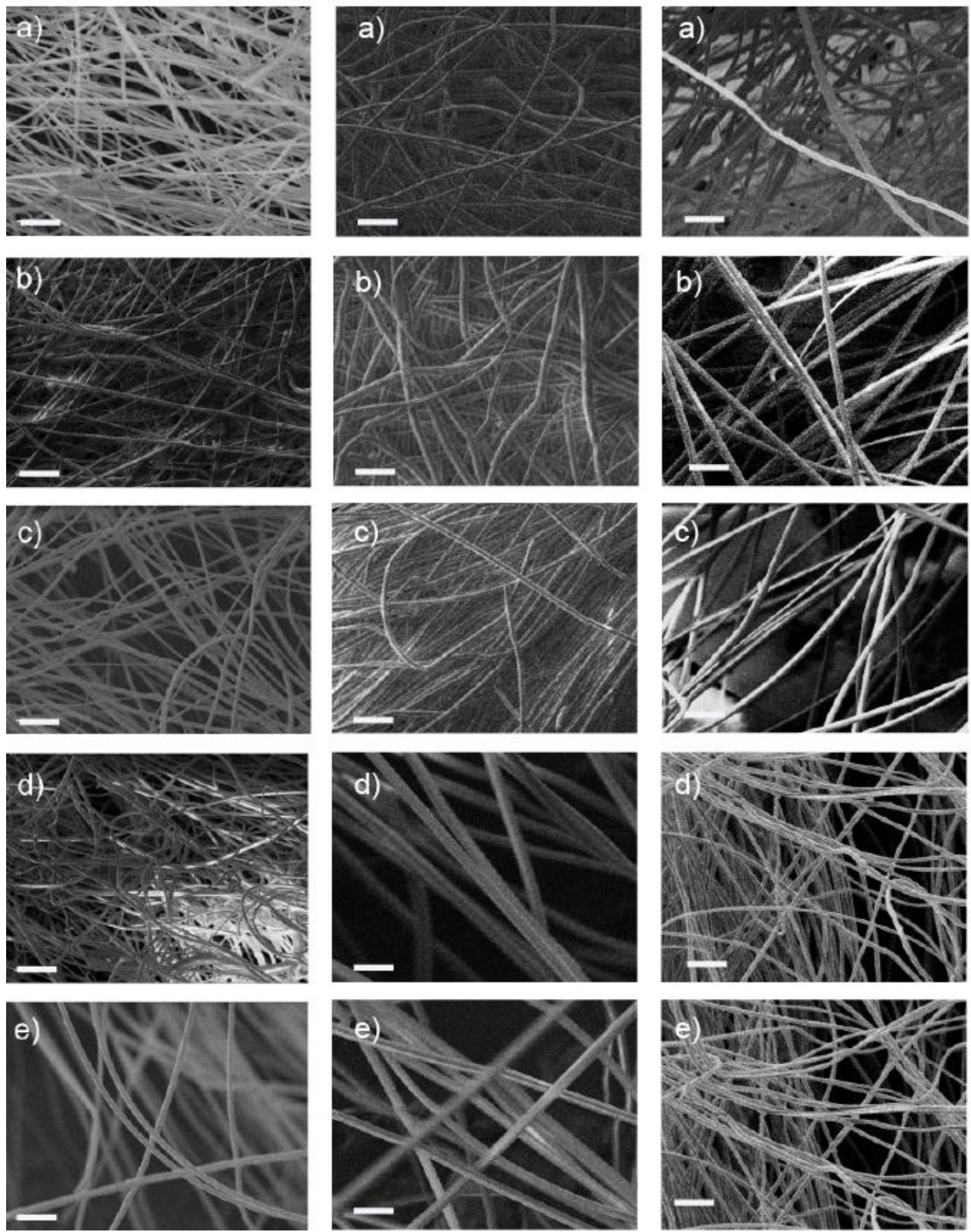

**Figure S8.** SEM images of (6,5)-SWNTs,pp-SWNTs and o-SWNTs samples. Left: a) (6,5)-SWNTs; b) MINT$_{(6,5)}$-**1**; c) MINT$_{(6,5)}$-**2**; d) (6,5) SWNTs-**1**; e) (6,5)-SWNTs-**2**. Center: a) pp-SWNTs; b) MINT$_{(pp)}$-**1**; c) MINT$_{(pp)}$-**2**; d) pp-SWNTs-**1**; e) pp-SWNTs-**2**. **Right:** a) o-SWNTs; b) MINT$_{(o)}$-**1**; c) MINT$_{(o)}$-**2**; d) o-SWNTs-**1**; e) o-SWNTs-**2**. Inset scale: 10 μm.

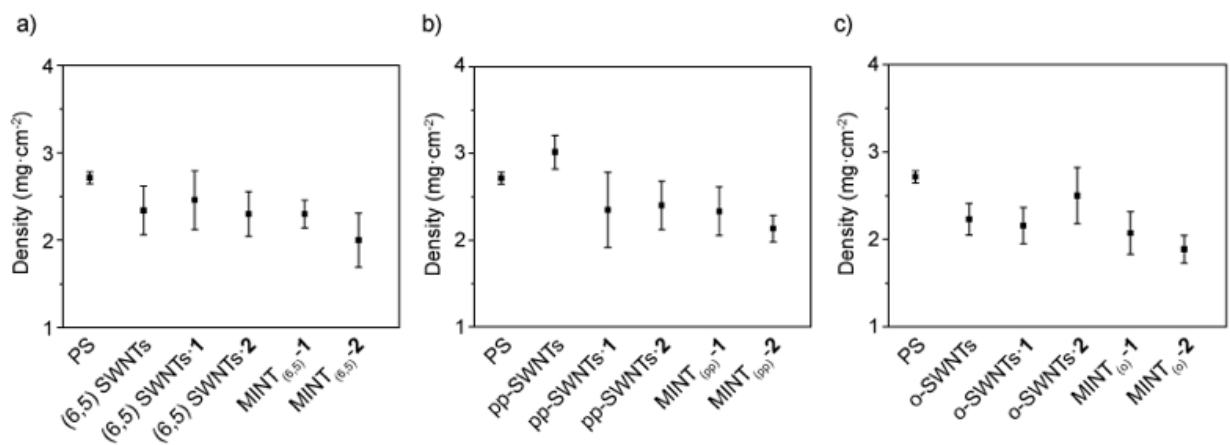

**Figure S9.** Fiber area density distribution: a) (6,5) SWNTs samples; b) pp-SWNTs samples and c) o-SWNTs. Mean of 7 measurements.

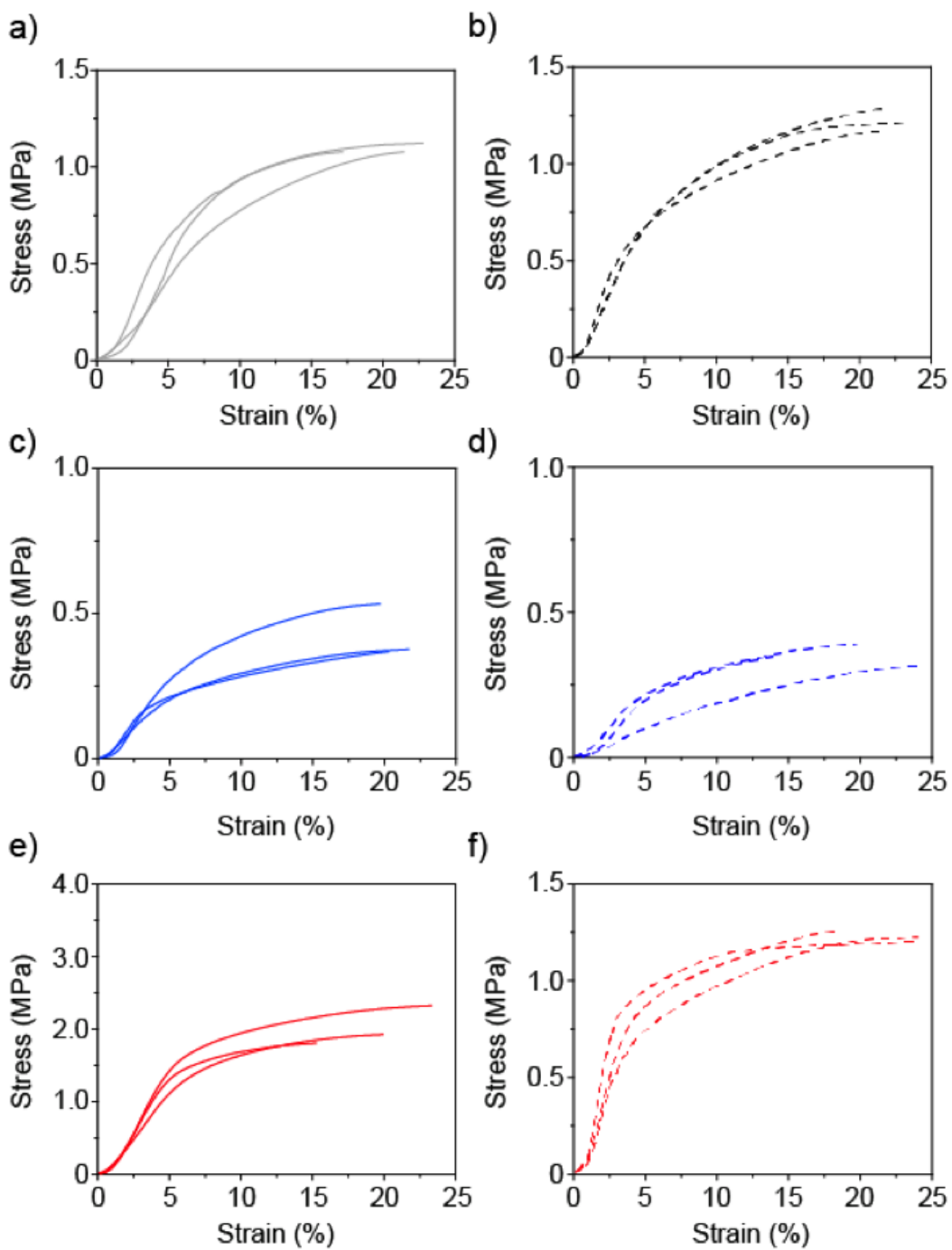

**Figure S10.** Stress/Strain curve of a) Polystyrene; b) (6,5)-SWNTs; c) (6,5)-SWNTs·**1**; d) (6,5)-SWNTs·**2**; e) MINT$_{(6,5)}$-**1** (red) f) MINT$_{(6,5)}$-**2**.

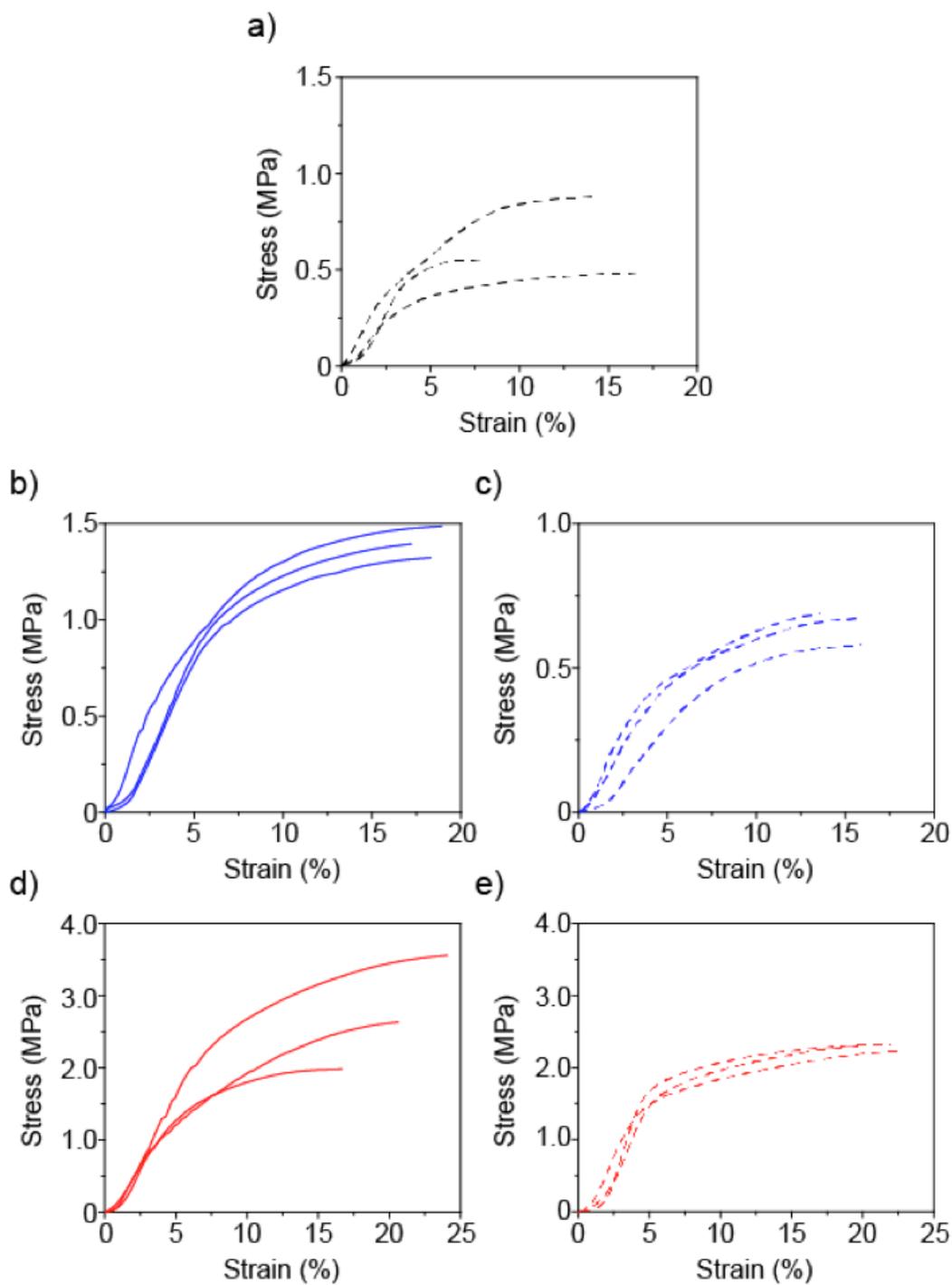

**Figure S11.** Stress/Strain curve of a) pp-SWNTs; b) pp-SWNTs·**1**; c) pp-SWNTs·**2**; d) MINT$_{(pp)}$-**1**; e) MINT$_{(pp)}$-**2**.

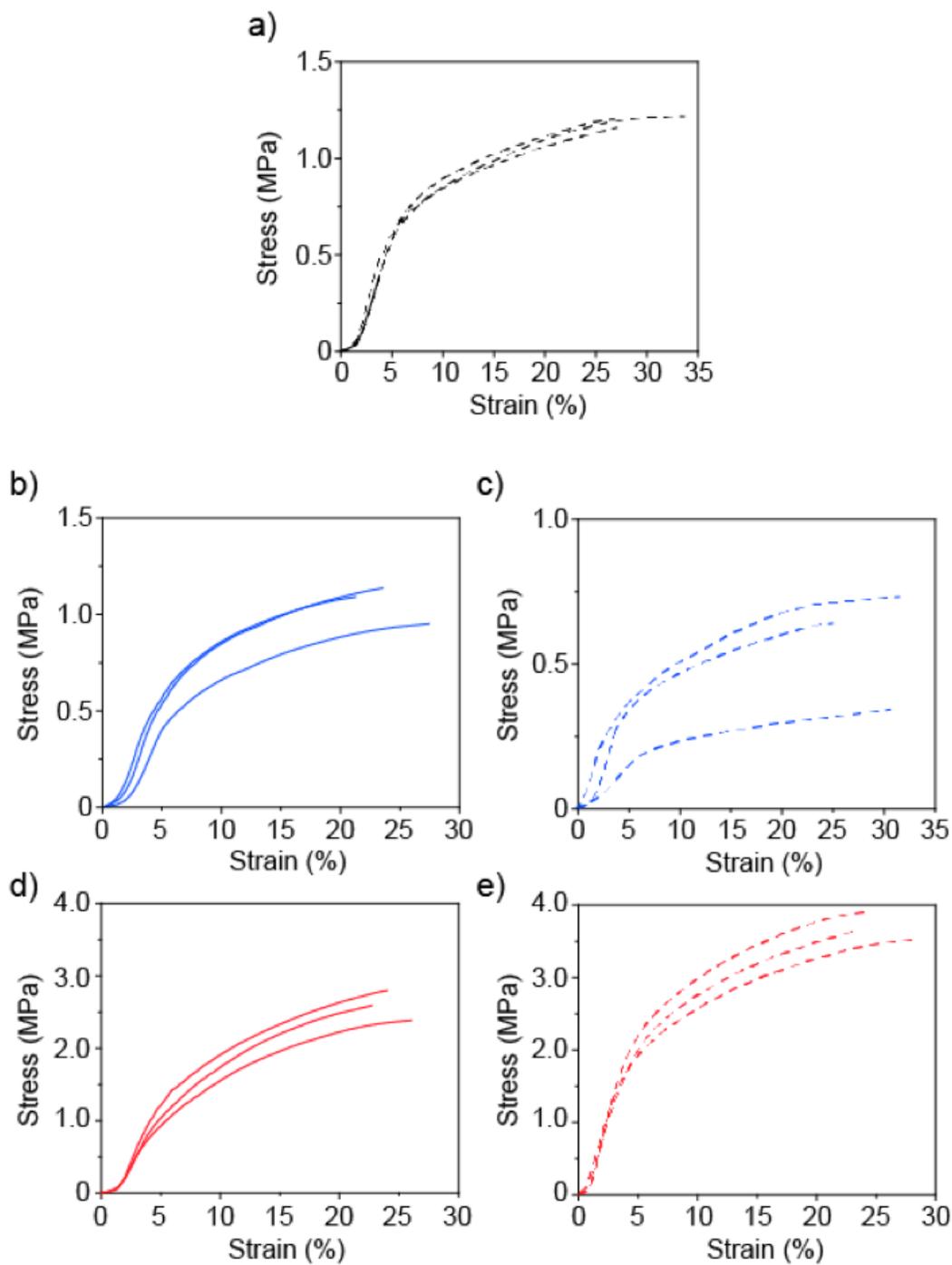

**Figure S12.** Stress/Strain curve of a) o-SWNTs; b) o-SWNTs·**1**; c) o-SWNTs·**2**; d) MINT$_{(o)}$-**1**; e) MINT$_{(o)}$-**2**.

**Table S1.** Young's Modulus and Tensile Strength in MPa of electrospun fibers.

| Sample | Interaction | Young's Modulus /MPa | Tensile Strength /MPa | Variation of YM respect PS /% | Variation of TS respect PS /% |
|---|---|---|---|---|---|
| Polystyrene (PS) | - | 15±1 | 1.09±0.03 | - | - |
| PS/ (6,5)-SWNTs | - | 18±1 | 1.26±0.06 | 20 | 16 |
| PS/ MINT$_{(6,5)}$-**1** | Mechanical bond | 32±6 | 2.0±0.3 | 110 | 80 |
| PS/**1**·(6,5)-SWNTs | Supramolecular | 7±1 | 0.39±0.03 | -53 | -60 |
| PS/ MINT$_{(6,5)}$-**2** | Mechanical bond | 21±7 | 1.18±0.08 | 40 | 8 |
| PS/**2**·(6,5)-SWNTs | Supramolecular | 6±1 | 0.37±0.05 | -60 | -66 |
| PS/ pp-SWNTs | - | 16±3 | 0.6±0.2 | 7 | -40 |
| PS/ MINT(pp)-**1** | Mechanical bond | 34±5 | 3.0±0.7 | 130 | 170 |
| PS/**1**·pp-SWNTs | Supramolecular | 23±4 | 1.4±0.1 | 50 | 30 |
| PS/ MINT$_{(pp)}$-**2** | Mechanical bond | 50±12 | 2.25±0.05 | 230 | 106 |
| PS/ **2**·pp-SWNTs | Supramolecular | 11±3 | 0.65±0.06 | -30 | -40 |
| PS/ o-SWNTs | - | 23±2 | 1.18±0.02 | 50 | 8 |
| PS/ MINT(o)-**1** | Mechanical bond | 35±6 | 2.6±0.3 | 130 | 140 |
| PS/ **1**· o-SWNTs | Supramolecular | 19±7 | 1.4±0.6 | 30 | 30 |
| PS/ MINT(o)-**2** | Mechanical bond | 59±7 | 3.7±0.2 | 290 | 240 |
| PS/ **2**· o-SWNTs | Supramolecular | 10±4 | 0.6±0.2 | -30 | -40 |

## Computational Methods

MD simulations were performed using AMBER 12 software package[1] for all calculations. Following the literature[2, 3] the AMBER99 force field[4] was used to model the SWNT, the polymer and the macrocycles **1** and **2**. For missing bonds, angle torsions, or Van Der Waals parameters not included in the AMBER99 force field, the values were transferred from the general AMBER force field (GAFF)[5]. The initial structures were minimized using two cycles of conjugated gradient minimization. During the initial cycle, the SWNT was kept in their starting conformation using a harmonic constrains with a force constant of 500 Kcal/mol-Å. This was followed by another minimization cycle where the SWNT was kept a harmonic restraint force constant of 10 Kcal/mol-Å. To allow a slow relaxation of the systems: SWNT-polymer, supramolecular **1** and **2**, MINT **1** and **2**, the minimized structures was heated slowly from 0 to 300 K during 0.5 ns (using a 2 fs time step) under of constant-pressure-constant-temperature conditions (NPT). Finally, we carried out 5.0 ns (using a 2 fs time step) of MD simulation in NPT ensemble to equilibrate the system at 300 k. The positions of all SWNT atoms were constrained with a weak 10 Kcal/mol-Å harmonic potential during all MD simulation. In the figure **S.13** we show the calculated potential energy for all studied systems. We observe how the energy increases during the first few ps, corresponding to our heating process form 0 k to 300k, then the energy remains constant and the equilibrium was considered to be reached. Analysis and visualization of MD trajectories were performed with VMD software.[6] In the figure **S.14** we show the initial and the final state of the molecular dynamic simulations conducted.

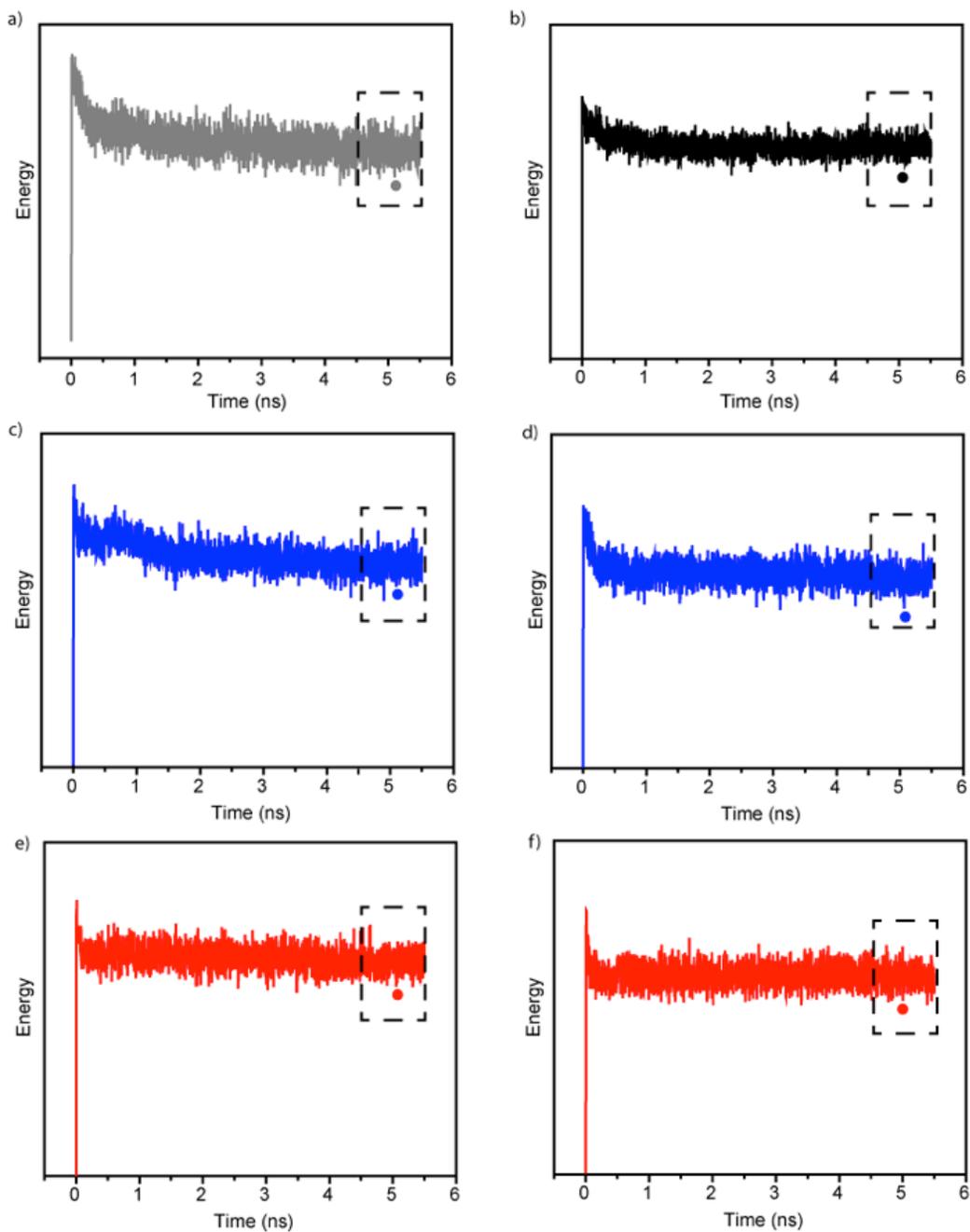

**Figure S13**. Stabilization energy of a) PS, and its composites with b) SWNTs, c) SWNT·**1,** d) SWNT·**2**, e) MINT-**1** and f) MINT-**2** during the 5.5 ns of MD simulations. The circles indicate the chosen structure for the figure 5. The dashed rectangles remark the chosen snap for the dihedral angle analysis.

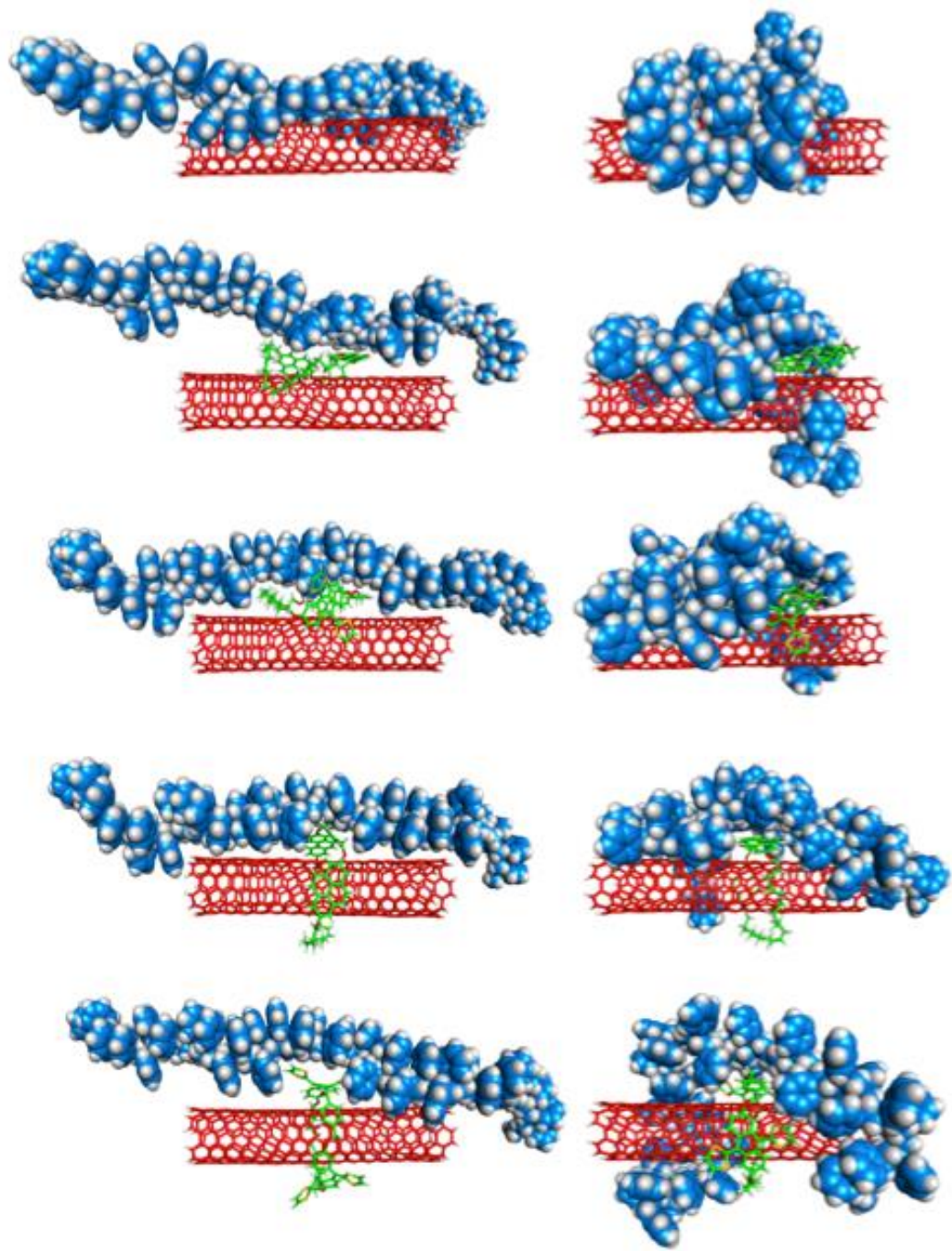

**Figure S14**. Initial and final state of the MD simulation for the five studied composites.